\documentclass[letterpaper, 10 pt, conference]{ieeeconf}  

\IEEEoverridecommandlockouts                              

\usepackage{graphics} 
\usepackage{epsfig} 
\usepackage{amsmath} 
\usepackage{amssymb}  
\usepackage{epstopdf}
\usepackage{dsfont}

\usepackage{framed}
\usepackage{scrextend}
\usepackage{subfigure}
\usepackage{cuted}

\usepackage{cases}
\usepackage{color}
\definecolor{rouge}{rgb}{1,0,0}
\definecolor{or}{rgb}{1,0.5,0}
\definecolor{bleu}{rgb}{0,0,1}
\definecolor{vert}{rgb}{0,1,0}

\newcommand{\forceindent}{\leavevmode{\parindent=1em\indent}}
\newcommand{\E}{\mathbb{E}}

\newcommand{\p}{\partial}
\newcommand{\x}{\mathrm{x}}

\newcommand{\ya}{\mathrm{y}}
\newcommand{\za}{\mathrm{z}}
\newcommand{\SOC}{\mathrm{SOC}}
\newcommand{\CSC}{\mathrm{CSC}}
\newcommand{\Cea}{\mathrm{C}_{e1}}
\newcommand{\Ceb}{\mathrm{C}_{e2}}

\newcommand{\T}{\mathrm{T}}
\newcommand{\Ts}{\mathrm{T}_s}
\newcommand{\tr}{\mathrm{T}}
\newcommand{\h}{\mathrm{h}}
\newcommand{\da}{\mathrm{d}}

\newcommand{\V}{\mathrm{V}}
\newcommand{\I}{\mathrm{I}}

\newcommand{\Pa}{\mathrm{P}}

\newcommand{\spt}{\hspace{-0.8cm}}

\newcommand{\spb}{\hspace{-0.3cm}}

\title{\LARGE \bf
Partition-based Unscented Kalman Filter for Reconfigurable Battery Pack State Estimation using an Electrochemical Model*
}

\author{Luis D. Couto and Michel Kinnaert$^{1}$
\thanks{*This research has been funded by Fonds pour la Formation à la Recherche dans l’Industrie et dans l’Agriculture (FRIA) of the FNRS.}
\thanks{This work is performed in the framework of the BATWAL project financed by the Walloon region (Belgium).}
\thanks{$^{1}$L. D. Couto and M. Kinnaert are with the School of Engineering of the  Universit\'e Libre de Bruxelles, B-1050 Bruxelles, Belgium
        {\tt\small lcoutome@ulb.ac.be, michel.kinnaert@ulb.ac.be}}%
}

\begin{document}

\maketitle
\thispagestyle{empty}
\pagestyle{empty}
\begin{abstract}
Accurate state estimation of large-scale lithium-ion battery packs 
is necessary for the advanced control of batteries, which could potentially increase their lifetime through e.g. reconfiguration. 
To tackle this problem, an \emph{enhanced} reduced-order electrochemical model is used here. 
This model allows considering a wider operating range and thermal coupling between cells, the latter turning out to be significant. 
The resulting nonlinear model is exploited for state estimation through unscented Kalman filters (UKF). 
A sensor network composed of one sensor node per battery cell is deployed. 
Each sensor node is equipped with a local UKF, which uses available local measurements together with additional information coming from neighboring sensor nodes. 
Such state estimation scheme gives rise to a partition-based unscented Kalman filter (PUKF). 
The method is validated on data from a detailed simulator for a battery pack comprised of six cells, with reconfiguration capabilities. 
The results show that the distributed approach outperforms the centralized one in terms of computation time at the expense of a very low increase of mean-square estimation error.
\end{abstract}

\section{INTRODUCTION}

Energy storage is a key point for a sustainable society based on environmentally friendly modes of transportation
and exploitation of renewable energy sources (for industry and housing). Among the different possibilities, lithium-ion
batteries are the most promising systems given their high energy and power density. Nevertheless, single battery cells are not able to provide the energy capacity or voltage required for large-scale applications, but they can be connected in series/parallel arrangements (battery pack) to cope with load specifications. Thus, battery packs might be comprised of hundreds or thousands of battery cells that need to be carefully monitored to ensure their safe operation.

The system in charge of battery supervision is the battery-management system (BMS). A BMS must be able to monitor internal state evolution, such as state-of-charge (SOC) for each cell ideally. Indeed, cell-to-cell variations arise due to the manufacturing process and the uneven operating conditions (e.g. temperature gradient) as well as ageing \cite{Xiong-2013}. 
Besides, battery pack configuration may change for equalization \cite{Zheng-2013} purposes 
or tolerance to faults. 
Therefore, our aim is to design a supervision system for a battery pack, which is able to provide an estimate for the internal state of each cell whatever the pack configuration. 
Besides, this supervision system should be distributed in order to avoid both reliability and communication issues linked to a centralized data fusion center and to ease scalability as well.

Battery pack state estimation requires appropriate battery pack models. Two approaches have been considered. 
The first one adopts a single cell to describe the entire battery pack \cite{Xiong-2013, Castano-2015}, 
which does not allow to estimate individual cell state. 
In the second approach, single cell models are linked to form a battery pack model that represents the battery pack behaviour \cite{Zhang-2016,Zhao-2015, Zheng-2013,Sun-2015}. 
Yet another option is to choose the weaker cells to describe the pack behaviour \cite{Zhong-2014,Hua-2015}. 
Most of the aforementioned battery pack modelling efforts are based on equivalent circuit models (ECM) of the cell. 
Some of them have been exploited for battery pack SOC estimation using different variants of Kalman filters \cite{Plett-2004c,Zhang-2016,Xiong-2013,Sun-2015,Zhong-2014}
or deterministic estimation approaches \cite{Zhao-2015, Zheng-2013, Hua-2015}.

As ECM parameters lack physical interpretability, 
these models are not appropriate to characterize battery state-of-health degradation by tracking their parameters. 
Since electrochemical model (EChM) parameters can directly be linked to the type of degradation (capacity fade, power fade), they are preferred for the present study. 
Such models have been used for the simulation of battery packs. 
More precisely, the Doyle-Fuller-Newman (DFN) model \cite{Newman-2004} has been exploited in \cite{Ci-2007, Kim-2009} to test battery pack dynamic reconfiguration. 
This model has been extended with thermal dynamics \cite{Wu-2013} and ageing \cite{Ashwin-2017} notably to gain insight in cell imbalances. 
Yet thermal coupling between cells \cite{Cordoba-2015} was not considered in these works despite the possibility of failure propagation through the entire battery pack due to a strong heat transfer \cite{Lamb-2015}. 
So far, no EChM was used for battery pack state estimation to the best of our knowledge.

All the aforementioned state observers for battery packs that account for cell-to-cell variations are based on ECMs and are centralized. 
For a large-scale battery pack equipped with a sensor network, distributed state estimation using so-called partitioned observers appears to be a suitable option. 
In this strategy, the system state space model is decomposed into a set of interacting subsystems. 
The state of each subsystem is estimated locally from the available local measurements, and possibly additional information obtained by data exchange with neighboring sensor nodes. 

Moving horizon, Luenberger-based and Kalman filter based partition observers have been developed for linear systems \cite{Farina-2010b, Riverso-2015, Farina-2016}. 
Yet, as EChMs are nonlinear, nonlinear versions of these filters must be sought. 
Here we resort to a partition-based unscented Kalman filter (PUKF) inspired by the work \cite{Singh-2014, Minot-2016} dealing with power system applications. 
While, in \cite{Singh-2014}, no coupling between subsystems is considered, in \cite{Minot-2016}, overlapping measurements and unequal subsystem dimensions impose specific features in the algorithm.

In this paper, we depart from previous work in the following ways:
\begin{itemize}
\item A simplified battery pack EChM accounting for thermal coupling between cells and exhibiting individual cell SOC as state variables is developed.
\item A partition-based unscented Kalman filter (PUKF) is designed on the basis of this model.
\item The PUKF is validated on a detailed battery pack simulator based on the DFN model and accounting for thermal dynamics coupling between cells and changes in battery cell interconnection.
\end{itemize}

The paper is organized as follows. The problem is stated in section \ref{sec:prob}. Two battery pack models are introduced in section \ref{sec:mod}, one for simulation and the other one for estimation. The state observer is designed in section \ref{sec:obs}. 
The results of the validation of the PUKF in simulation are given in section \ref{sec:sim}.

{\bf Notation:}
$\mathrm{diag}\{X\}$ denotes a diagonal matrix with the entries of $X$ on the diagonal. 
$\mathrm{col}\{ x_1,\ldots,x_n \}$ is a vector obtained by stacking the vectors $x_1,\ldots,x_n$. 
Boldface letters ${\bf A}$ denote system (collective) variables and normal letters $A$ denote subsystems (partitioned) variables.
$1_{n}$ is a row vector of size $n$ with all entries equal to one.
$0_{m\times n} \in \mathbb{R}^{m \times n}$ is a matrix with all entries equal to zero.


\vspace{-0.08cm}
\section{PROBLEM STATEMENT} \label{sec:prob}
\vspace{-0.08cm}

To be able to state the problem, both the battery pack configurations and the associated sensor network are presented.

\begin{figure}[!htb]
\centering
\subfigure[]{\label{fig:sw}\includegraphics[scale=0.46]{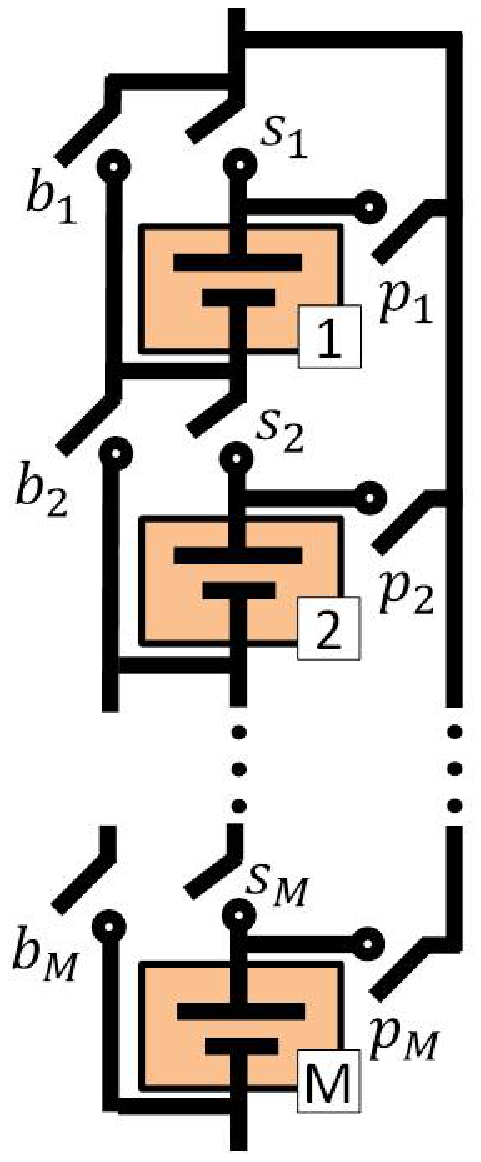}}
\subfigure[]{\label{fig:layout}\includegraphics[scale=0.46]{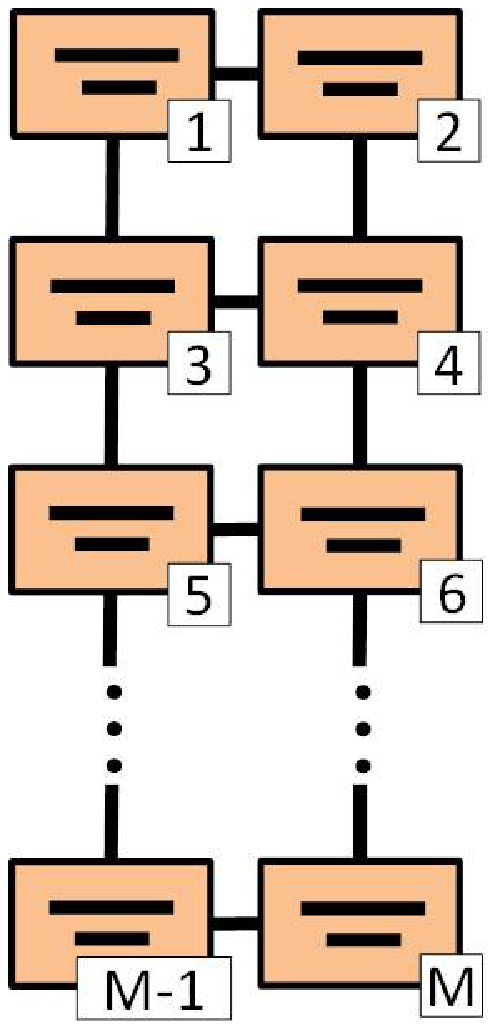}}
\subfigure[]{\label{fig:reconf}\includegraphics[scale=0.46]{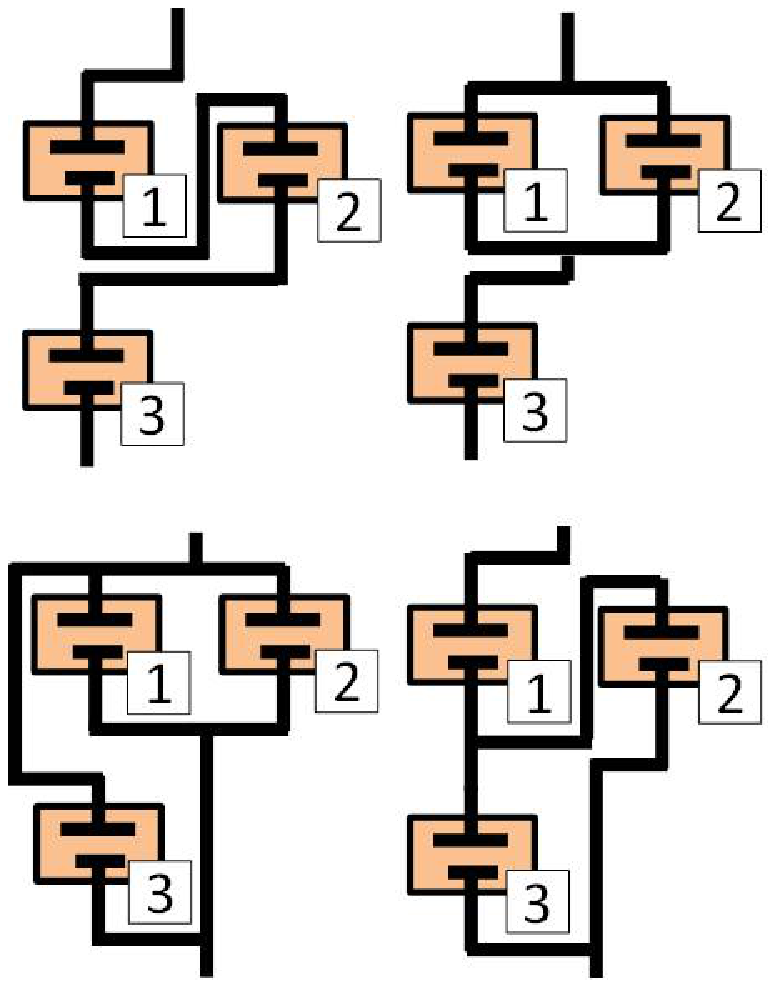}}
\vspace{-0.2cm}
\caption{Reconfigurable battery pack, namely a) switches deployment, and examples of b) battery cells physical layout for $N_p = 2$, and c) all feasible configurations for $M = 3$ battery cells.}
\vspace{-0.4cm}
\label{fig:conf}
\end{figure}

A reconfigurable battery pack made of series/parallel arrangements is considered, because it allows synthesizing different capacities/voltages. It consists of a battery array made of $M$ battery cells with the associated switches depicted in Figure \ref{fig:sw} \cite{Kim-2009,Kim-2012x}.
It has been shown \cite{Kim-2012x} that equipping each battery cell with three switches suffices to allow for reaching any configuration ranging from all cells in series to all cells in parallel.
Switches $s_i$, $p_i$ and $b_i$, $i = 1,\ldots,M$ denote respectively series, parallel and bypass, although bypassing operations are omitted in this contribution. 
Assuming that all the switches $s_i$, $p_i$ and $b_i$ are opened at the begining, 
battery cells $i$ and $i+1$ are put in series when switches $(s_i, s_{i+1})$ are closed, and in parallel when switches $(p_i, p_{i+1})$ and $(b_i, b_{i+1})$ are closed. 

{\bf Remark:} {\it The products on the market are typically made of a series arrangement of $N_s$ groups of cells, each group being made of $N_p$ cells in parallel (with $M = N_s \times N_p$). 
This topology is denoted as series-parallel (SP). The SP topology is expected to be robust against individual cell faults since it avoids voltage drops upon cell failure. 
However, heterogeneous configurations can be considered as well, where the number of cells in parallel in each group is different, say $N_{p_i}$ for $i = 1,\ldots,N_s$ (with $M = \sum_{i=1}^{N_s} N_{p_i})$. 
Heterogeneous configurations can be more cost-effective \cite{Jin-2012} and they allow to conveniently serialize groups of battery cells to cope with single or even multiple loads and parallelize other groups to naturally balance charge, manage ageing and facilitate charging.}

With regard to sensing, the sensor network is composed of one sensor node per battery cell. Each sensor node measures the cell voltage, current and surface temperature. Communication only takes place between neighboring nodes, namely nodes associated to cells that are physically placed beside each other as illustrated in Figure \ref{fig:layout}. 
Notice that the trend towards BMS based on highly instrumented battery pack can be observed in some references \cite{REC}.

Our aim is thus to estimate the internal state of each battery cell via a distributed state observer despite the thermal coupling and the variety of electrical configurations of the pack. 
The validation of the estimation scheme is performed on a detailed battery pack simulator which is described in the next section.


\section{MODELING}	\label{sec:mod}

Two lithium-ion EChMs are considered in this contribution, which differ in their complexity. On the one hand, the well known DFN battery cell model \cite{Newman-2004} 
is used as a basis for developing a reconfigurable battery pack simulator. 
On the other hand, a reduced-order EChM based on \cite{Couto-2016} is used for designing a distributed state observer. 
Both models represent the dynamic behaviour of in-pack battery cells, which have to be properly interconnected to meet energy capacity and/or voltage requirements. Such interconnection is considered to be time-varying due to topological dynamic reconfiguration. 


\subsection{Battery pack simulator} \label{sec:recongBP}

The model of the battery pack simulator builds up from the basic units, i.e. the battery cells. Each cell is represented by a DFN model \cite{Newman-2004}, 
that describes the electrochemical processes within the cell. 
This model resort to porous electrode and concentrated solution theories to describe solid and solution phases of the battery cell, respectively. It consists of a set of coupled nonlinear algebraic and partial differential equations (PDEs) that should be solved at each time instant. Each model implements the solid-phase diffusion equations through a third-order Pad\'e approximation and discretizes all the other equations through a central difference method. 
The battery chemistry is the standard graphite/LCO, whose parameters are publicly available in the DUALFOIL model \cite{Newman-1998}. 

The individual cell models are interconnected both thermally and electrically within series/parallel arrangements. We successively address the modelling of a possibly time-varying topology, the thermal interconnection and the electrical interconnection below.

\subsubsection{Configuration of the Battery Pack}

Notice in Figure \ref{fig:sw} the sequential property of in-pack battery cells \cite{He-2017}, i.e. the monotonically increasing cells indices. This property forces neighboring cells to electrically pair with each other. 
This aspect and the SP topology limits the reconfiguration combinatorial problem to $2^{M-1}$ feasible configurations. 
By gathering together all possible battery pack configurations, a switched system can be abstracted. 
Thus, the configuration at time instant $k$ can be described by a switching signal $\sigma(k)$ that takes values in the set $\mathcal{P} = \{ 1, 2, \ldots, 2^{M-1} \}$. 
Without loss of generality, let us consider 
a simple example with $M = 3$ battery cells 
to illustrate the proposed dynamic reconfiguration.
In this case, there are $2^2$ feasible configurations, $\mathcal{P} = \{ 1,\ldots,4 \}$. 
They are depicted in Figure \ref{fig:reconf} where the physical layout displayed in Figure \ref{fig:layout} was kept.

\subsubsection{Thermal Interconnection}

The source of dynamic coupling among in-pack battery cells is heat exchange. 
Only neighboring cells can exchange heat, meaning that the thermal interconnection topology is position-dependent. 

As the sensor network is also associated to the cell position and communication only arises between neighboring nodes (namely neighboring battery cells), the adjancency matrix $Ad$ of the sensor network will be used to characterize thermal coupling below. 
We will also denote by $\mathcal{N}_i$ the neighborhood of node $i$.

From a thermal energy balance of the $i$-th battery cell \cite{Bernardi-1985}
and its reformulation in terms of two-state thermal model \cite{Lin-2014} 
with $[\T_c, \T_s]$ as \emph{core} and \emph{surface} temperature, respectively, 
the following continuous-time model for $\T_c$ is obtained\footnote{Continuous-time with time variable $t$ is adopted only in this section since subsystems interconnection is stated in terms of the DFN model, which is usually presented in this time domain.} 
\begin{eqnarray}
\rho_{c,i} C_{pc,i} \dot{\T}_{c,i}(t)
\hspace{-0.2cm}&= 
\hspace{-0.2cm}& - \mathrm{J}_i(t)\V_i(t) - k_{c,i} \left( \T_{c,i}(t) - \T_{s,i}(t) \right) \nonumber \\
&& \hspace{-2.4cm} + \left(\hspace{-0.1cm} \left. \int \limits_{0}^{L_B} \hspace{-0.1cm}F a_s j_n(x,t) \left(\hspace{-0.1cm} U(x,t) - \T_{c,i}(t) \frac{\partial U}{\partial T}(x,t) \hspace{-0.1cm}\right) \hspace{-0.1cm}dx \hspace{-0.1cm}\right) \right\vert_i \label{eq:T2state_1}
\end{eqnarray}

where $[\mathrm{J}_{i}, \mathrm{V}_{i}]$ are battery cell input current and output voltage, respectively, 
$c$ subindex refers to \emph{core} variables, 
$\rho$ and $C_p$ are the density and specific heat, respectively, 
$k_c$ is the thermal conductivity, 
$F$ is the Faraday's constant, $a_{s}$ is the specific interfacial area, and $j_{n}$ and $U$ are the pore-wall molar flux and open-circuit voltage, respectively. 
The latter two variables are $x$-position dependent, with $x$ defining the axis along the cell thickness $L_B$.

For its part, a continuous-time model for $\T_s$ accounting for the heat exchange between the $i$-th cell and its neighboring $j$ cells 
can be written in the following matrix form
\begin{equation}
\begin{array}{lcl}
E \dot{{\bf T}}_{{\bf s}}(t)
\hspace{-0.2cm}&= 
\hspace{-0.2cm}&( A + Ad K_e ) {\bf T_s}(t) + K_c {\bf T}_c(t) \\
&&\hspace{0.8cm}	+ 1^\tr_{M} \h T_{\infty} - R_{\mathrm{IC}} {\bf I}(t)^2
\label{eq:statematT}
\end{array}
\end{equation}

where ${\bf T_s}(t) = \left[ \T_{s,1}(t), \ldots, \T_{s,M}(t) \right]^\tr$ (similar for ${\bf T_c}(t)$ and ${\bf I}(t)$), 
$E = \mathrm{diag}\{ \rho_{s,1} C_{ps,1}, \ldots, \rho_{s,M} C_{ps,M} \}$ with $s$ subindex referring to \emph{surface} variables, 
$A = - \mathrm{diag}\{ k_{g,1}, \ldots, k_{g,M} \}$ with $k_{g,i} = \h + k_{c,i} + \sum_{j \in \mathcal{N}_i} k_{e,j}$, in which 
$\h$, $T_\infty$ and $k_{e,j}$ are the heat transfer coefficient, ambient temperature (assumed as constant) and thermal conductivity with the adjacent $j$-th cell, respectively. 
$Ad$ is the adjacency matrix of the sensor network defined by Boolean entries such that 
$Ad(i,j) \neq 0$ if $j \in \mathcal{N}_i, i \neq j$ 
and 
$K_e = \mathrm{diag}\{ k_{e,1}, \ldots, k_{e,M} \}$ (similar for $K_c$ and $R_{\mathrm{IC}}$). 
The last term on the right-hand side of Eq. (\ref{eq:statematT}) accounts for the Joule heating due to inter-cell connection resistances $R_{\mathrm{IC}}$ \cite{Wu-2013}.

\subsubsection{Electrical Interconnection}

Besides thermal dynamic coupling, in-pack battery cells are also algebraically coupled due to electric charge exchange. 
The electrical interconnection topology is switching-dependent (Figure \ref{fig:sw}), in contrast with the thermal one. 
From Kirchhoff's laws, the following equation is obtained
\begin{equation}
F_\sigma {\bf J}(t)
= 
G_\sigma {\bf V}(t) + H_\sigma \I(t) \label{eq:statematI}
\end{equation}

where ${\bf J}(t) = \left[ \mathrm{J}_{1}(t), \ldots, \mathrm{J}_{M}(t) \right]^\tr$ (similar for ${\bf V}(t)$), 
$\mathrm{I}(t)$ is the battery pack input current, 
$F_\sigma$, $G_\sigma$ and $H_\sigma$ are two block diagonal matrices and a column vector, respectively, where subindex $\sigma$ denotes their dependency with the switching signal.
The entries of the matrices for cells in series take the form:
\begin{equation}
F_i^s = 1,
\hspace{0.1cm}
G_i^s = 0, 
\hspace{0.1cm}
H_i^s = 1
\end{equation}

while for an interconnection of $n_p$ cells in parallel, the associated blocks take the form
\begin{equation}
\hspace{-0.2cm}
F_i^p 
\hspace{-0.1cm}
=
\hspace{-0.1cm}
\left[
\hspace{-0.2cm}
\begin{array}{ccccc}
1	\spb& 1	\spb& \cdots 	\spb& 1 \spb& 1 \\
-R_{\mathrm{IC},1}	\spb& R_{\mathrm{IC},1}	\spb& \cdots		\spb& 0 \spb& 0\\
\vdots \spb& \vdots \spb& \ddots 	\spb& \vdots \spb& \vdots \\
0 	\spb& 0	\spb& \cdots \spb& R_{\mathrm{IC},n_{p}-1} \spb&0 \\
0 	\spb& 0	\spb& \cdots 	\spb& -R_{\mathrm{IC},n_{p}}	\spb&R_{\mathrm{IC},n_{p}}
\end{array}
\hspace{-0.2cm}
\right]
\end{equation}

\begin{equation}
\hspace{-0.2cm}
G_i^p 
\hspace{-0.1cm}
=
\hspace{-0.1cm}
\left[
\hspace{-0.2cm}
\begin{array}{ccccc}
0	\spb& 0	\spb& \cdots 	\spb& 0 \spb& 0 \\
1	\spb& -1	\spb& \cdots		\spb& 0 \spb& 0\\
\vdots \spb& \vdots \spb& \ddots 	\spb& \vdots \spb& \vdots \\
0 	\spb& 0	\spb& \cdots \spb& -1 \spb&0 \\
0 	\spb& 0	\spb& \cdots 	\spb& 1	\spb&-1
\end{array}
\hspace{-0.2cm}
\right]
\hspace{-0.1cm}
,
H_i^p 
\hspace{-0.1cm}
=
\hspace{-0.1cm}
\left[
\hspace{-0.2cm}
\begin{array}{c}
1 \\
0 \\
\vdots \\
0 \\
0 
\end{array}
\hspace{-0.2cm}
\right]
\end{equation}

where $F_i^p$, $G_i^p \in \mathbb{R}^{n_p\times n_p}$ 
and $H_i^p \in \mathbb{R}^{n_p\times 1}$.


\subsection{Reduced-Order Electrochemical Model}

A reduced-order distributed model of a reconfigurable battery pack is now developed, which is suitable for real time state estimation. 
In this model, each cell is associated to a state vector $\mathrm{x}_i$, $i = 1,\ldots,M$. 
The cell state vectors are non-overlapping, and the dynamics of each battery cell is described as:
\begin{eqnarray}
\hspace{-0.4cm} \mathrm{x}_i (k+1) \hspace{-0.3cm}&= \hspace{-0.4cm}& 
\sum\limits_{j \in \mathcal{N}_i} \hspace{-0.1cm} A_{ij} \mathrm{x}_j (k) 
\hspace{-0.1cm} + \hspace{-0.1cm} f_i ( \mathrm{z}_i (k), \mathrm{x}_i (k), \mathrm{y}_i (k), w_i(k) )   \label{eq:distmodel} \\
\hspace{-0.4cm} \mathrm{y}_i (k) \hspace{-0.3cm}&= \hspace{-0.3cm}& h_i \left( \mathrm{z}_i(k), \mathrm{x}_i(k) v_i(k) \right) \label{eq:distout}
\end{eqnarray}

where $\mathrm{x}_i (k) \in \mathbb{R}^{n_i}$ is the state, 
$\mathrm{y}_i (k) \in \mathbb{R}^{m_i}$ is the local measured output and 
$\mathrm{z}_i (k) \in \mathbb{R}^{p_i}$ is the local input, 
$w_i (k) \in \mathbb{R}^{n_i}$ and $v_i (k) \in \mathbb{R}^{m_i}$ are respectively process and measurement zero-mean Gaussian noise sequences, which are mutually uncorrelated such that
\begin{equation}
\E\left[ \left[ \begin{array}{c}
w_i(k) \\
v_i(k) \end{array} \right] 
\left[ w_i (m)^{\mathrm{T}} v_i(m)^{\mathrm{T}} \right] \right]
=
\left[ \begin{array}{cc}
Q_i & 0 \\
0 & R_i
\end{array} \right] \delta_{km} \label{eq:disturbance}
\end{equation}

where $\E[\cdot]$ is the expectation operator, $\delta_{km}$ is the Kronecker delta, $Q_i$ and $R_i$ are the process and measurement noise variances, respectively.

The matrices and nonlinear functions in \eqref{eq:distmodel}, \eqref{eq:distout} were determined by extending the equivalent-hydraulic model (EHM) derived in \cite{Couto-2016} by incorporating both electrolyte and thermal dynamics. This enhanced EHM is denoted as $e$EHMT below. 
Such an $e$EHMT covers a wider operating range than the original EHM \cite{Couto-2016}, including higher C-rates ($\geq 1 C$) and thermal gradients. 
Its derivation is detailed in the Appendix. It was inspired by \cite{Rahn-2013ch4} for what regards the electrolyte dynamics. 
As far as the thermal dynamics is concerned, it was deduced from Eqs. (\ref{eq:T2state_1}) and (\ref{eq:statematT}) by considering an uniform 
pore-wall molar flux and an average open-circuit voltage along the cell thickness. 

The cell state vector has the form:

\vspace{-0.5cm}
{\small
$$\x_i(k) = \left[ \SOC_i(k), \CSC_i(k), \mathrm{C}_{e1,i}(k), \mathrm{C}_{e2,i}(k), \T_{c,i}(k), \T_{s,i}(k) \right]^{\tr}$$
}
\vspace{-0.6cm}

where $\SOC$ is the state-of-charge, $\CSC$ is the critical surface concentration, $\Cea$ and $\Ceb$ 
characterize the electrolyte diffusion. They can be seen as the equivalent of $\SOC$ and $\CSC$, respectively, in the electrolyte diffusion model. 
Besides, $\T_c$ and $\Ts$ are the core and surface temperature, respectively, $\za_i(k) = \mathrm{J}_i(k)$ is the local input current and $\ya_i(k) = \left[ \mathrm{V}_i(k), \mathrm{T}_{s,i}(k) \right]^\tr$ is the local measured output, namely the voltage and the surface temperature for the $i$-th battery cell. The state transition matrices are given as
\begin{equation}
A_{ii} 
\hspace{-0.1cm}
= 
\hspace{-0.1cm}
\left[
\hspace{-0.1cm}
\begin{array}{cccccc}
1 \hspace{-0.2cm}&0 \hspace{-0.2cm}&0 \hspace{-0.2cm}&0 \hspace{-0.2cm}&0 \hspace{-0.2cm}&0 \\
\frac{\mathrm{g}_{s,i}}{b_{s,i}} \hspace{-0.2cm}& 1 - \frac{\mathrm{g}_{s,i}}{b_{s,i}} \hspace{-0.2cm}&0 \hspace{-0.2cm}&0 \hspace{-0.2cm}&0 \hspace{-0.2cm}&0 \\
0 \hspace{-0.2cm}&0 \hspace{-0.2cm}&1 \hspace{-0.2cm}&0 \hspace{-0.2cm}&0 \hspace{-0.2cm}&0 \\
0 \hspace{-0.2cm}&0 \hspace{-0.2cm}&\frac{\mathrm{g}_{e,i}}{b_{e,i}} \hspace{-0.2cm}&1 - \frac{\mathrm{g}_{e,i}}{b_{e,i}} \hspace{-0.2cm}&0 \hspace{-0.2cm}&0 \\
0 \hspace{-0.2cm}&0 \hspace{-0.2cm}&0 \hspace{-0.2cm}&0 \hspace{-0.2cm}&1-k_{c,i} \hspace{-0.2cm}&k_{c,i} \\
0 \hspace{-0.2cm}&0 \hspace{-0.2cm}&0 \hspace{-0.2cm}&0 \hspace{-0.2cm}&k_{c,i} \hspace{-0.2cm}&1-k_i \\
\end{array}
\hspace{-0.1cm}
\right]
\label{eq:statemat}
\end{equation}

\begin{equation}
A_{ij} = 
\left[
\begin{array}{cccccc}
0_{5\times5} \hspace{-0.2cm}& 0_{5\times1}\hspace{-0.2cm} \\
0_{1\times5} \hspace{-0.2cm}&k_{c,ij}\hspace{-0.2cm} \\
\end{array}
\right]
\label{eq:Aij}
\end{equation}

where $A_{ii},A_{ij} \in \mathbb{R}^{n_i \times n_i}$, 
$\mathrm{g}_s = \frac{D_s}{R_s^2}$ with $D_s$ and $R_s$ as the solid-phase diffusion coefficient and particle radius, respectively, $b_s = \beta_s (1 - \beta_s)$ with $\beta_s \in (0,1)$, 
electrolyte variables $\mathrm{g}_e$ and $b_e$ are derived in the Appendix, 
and $k_i = \h + \sum_{j \in \mathcal{N}_i} k_{c,j}$. 
The nonlinear functions $f_i$ and $h_i$ are respectively given by:
\begin{equation}
f_{i} 
\hspace{-0.1cm}
= 
\hspace{-0.1cm}
\left[
\begin{array}{c}
\hspace{-0.2cm} - \gamma_s \hspace{-0.2cm}\\
\hspace{-0.2cm} \frac{\gamma_s}{1 - \beta_{s,i}} \hspace{-0.2cm}\\
\hspace{-0.2cm} -\gamma_e \hspace{-0.2cm}\\
\hspace{-0.2cm} \frac{\gamma_e}{1 - \beta_{e,i}} \hspace{-0.2cm}\\
\hspace{-0.2cm} \left( \Delta U_{b,i}^{\pm} - \V_i(k) \right)
- \Delta \frac{\partial{U_{b,i}^{\pm}}}{\partial{\T_{c,i}}} \T_{c,i}(k) \hspace{-0.1cm}\\
\hspace{-0.2cm} R_{\mathrm{IC},i}\mathrm{J}_i(k) \hspace{-0.2cm}\\
\end{array}
\hspace{-0.1cm}
\right] 
\hspace{-0.1cm}
\mathrm{J}_i(k) 
\hspace{-0.1cm}
+
\hspace{-0.1cm}
\left[
\begin{array}{c}
\hspace{-0.2cm} 0 \\
\hspace{-0.2cm} 0 \\
\hspace{-0.2cm} 0 \\
\hspace{-0.2cm} 0 \\
\hspace{-0.2cm} 0 \\
\hspace{-0.2cm} \h \\
\end{array}
\hspace{-0.1cm}
\right]
\hspace{-0.1cm}
\T_{\infty}
\label{eq:Bm}
\end{equation}
\begin{equation}
h_i
=
\left[
\begin{array}{c}
\Delta U_{s,i}^{\pm} + \Delta \eta^{\pm}_{s,i}  - \Delta\phi_{e,i} - R_{c,i} \mathrm{J}_i(k) \\
\mathrm{T}_{s,i}(k)
\end{array}
\right]		\label{eq:outputi}
\end{equation}

where 
$\gamma_s = \frac{3}{R_s c_{s,\mathrm{max}}}\frac{1}{F a_s L^-}$ with $c_{s,\mathrm{max}}$ as the maximum solid concentration and $L^-$ as the negative electrode thickness, 
$\gamma_e$ is derived in the Appendix and
$R_c$ is the solid-electrolyte interface film resistance, 
$\Delta(\xi)^{\pm}$ is the difference between functions $\xi^{+}$ and $\xi^{-}$, 
and functions $U^{\pm}_b$, $\frac{\partial{U}^{\pm}_b}{\partial{\T}_{c}}$, $U^{\pm}_s$, $\eta^{\pm}_s$, and $\Delta \phi_e$ are given in 
the Appendix.

%





Note that the coupling between cell dynamics is only due to thermal effects (see matrix $A_{ij}$). 
Besides, since thermal dependent parameters that follow the Arrhenius equation introduce significant nonlinearities into the functions $f_i$ and $h_i$, a partitioned unscented Kalman filter (UKF) is developed below instead of a partitioned extended Kalman filter relying on model linearization.

{\bf Remark:} {\it For the sake of comparison, a centralized UKF will be considered in the simulation section. 
This UKF is based on the global model obtained from \eqref{eq:distmodel},\eqref{eq:distout} by aggregating the state vectors as ${\bf x}(k) = \mathrm{col} \left\{ \mathrm{x}_1(k), \ldots, \mathrm{x}_M(k) \right\}$ 
and similarly for ${\bf y}(k), {\bf z}(k), {\bf w}(k)$ and ${\bf v}(k)$.}

\section{PARTITION-BASED STATE OBSERVER}	\label{sec:obs}

In order to estimate the state of system (\ref{eq:distmodel}),(\ref{eq:distout}), a partition-based unscented Kalman filter (PUKF) is developed in this section. 
In this algorithm, node $i$ of the sensor network estimates its local state $\hat{\x}_i$ and covariance $\Pa_i$, which are thus associated to the $i$-th battery cell ($i = 1, \ldots, M$).

The algorithm of Table \ref{tab:ukf} is deduced from \cite{Wan-2002}. 
The notation used for sigma point generation, namely $\sqrt{\Pa_{j,k-1}^a}$ stands for $\sqrt{(\Pa_{j,k-1}^a)_l}$, $l = 1,\ldots,n_j$ where $(\cdot)_l$ is the $l$-th column of the matrix square root. 
Due to the coupling in the state equation, the nodes have to exchange their state estimate and covariance as seen in Eqs. \eqref{eq:sigmpoints},\eqref{eq:timeup1}. 
Similarly as in \cite{Riverso-2015, Farina-2016}, the distributed nature of the filter is ensured by neglecting the off-diagonal terms in the state covariance matrix, as compared to a centralized approach. 
Yet to guarantee that the estimate is consistent (or conservative) as defined in \cite{Uhlmann-2003}, the inequality 
\begin{equation}
\mathrm{diag} \{ \Pa_1, \ldots, \Pa_M \} - \Pa_c \geq 0 \label{eq:Pi}
\end{equation}

should be ensured. In Eq. \eqref{eq:Pi}, $\Pa_c$ stands for the covariance matrix of the state estimation error in the centralized framework. 
Such a condition is also enforced in \cite{Farina-2016} in the framework of partition-based distributed Kalman filtering. 
To ensure this goal, an off-line approach is used here. 
First, $\Pa_c$ is computed for a centralized UKF with a standard value for $\alpha$, namely $\alpha = 10^{-2}$ and a charge/discharge profile covering a wide range of operating conditions (see next section). 
Next, parameter $\alpha$ is adjusted in the PUKF in order to ensure fulfilment of Eq. \eqref{eq:Pi} for the same data set. 
Such an off-line approach has been also proposed in \cite{Sakai-2010} in a different context.



\begin{table}[!htb]
\caption{Local unscented Kalman filter for the $i$-th subsystem\textsuperscript{\textdagger}.}
\vspace{-0.2cm}
\centering

{\small
\begin{tabular}{l}
\hline \vspace{-0.2cm}
\\
Initialization: for $k = 0$, set \\
\forceindent $\hat{\mathrm{x}}_{i,0} = \E [ \mathrm{x}_{i,0} ]$, \hspace{0.1cm}
$\Pa_{\x,i,0} = \E[ (\mathrm{x}_{i,0} - \hat{\mathrm{x}}_{i,0}) (\mathrm{x}_{i,0} - \hat{\mathrm{x}}_{i,0})^{\mathrm{T}} ]$ \\
\forceindent $\hat{\mathrm{x}}_{i,0}^a = \E[ \mathrm{x}_{i,0}^a ] = [ \hat{\mathrm{x}}_{i,0} \ \ 0 \ \ 0 ]^{\mathrm{T}}$ \\ 
\forceindent $\Pa_{i,0}^a = \E[ (\x_{i,0}^a - \hat{\x}_{i,0}^a)(\x_{i,0}^a - \hat{\x}_{i,0}^a)^\tr ] = \mathrm{diag}(\Pa_{\x,i,0}, Q_i, R_i)$
\vspace{0.1cm}\\
For $k = 1, 2, \ldots$ compute \\
Sigma points: \vspace{-0.2cm}\\
\vbox{
\begin{eqnarray}
\hspace{-0.3cm}\mathcal{X}^a_{j,k-1} &\hspace{-0.3cm}= &\hspace{-0.3cm} \left[ \hat{\mathrm{x}}^a_{j,k-1} \ \ \hat{\mathrm{x}}^a_{j,k-1}\hspace{-0.1cm} +\hspace{-0.1cm} \gamma_j \sqrt{\mathrm{P}^a_{j,k-1}} \ \ \hat{\mathrm{x}}^a_{j,k-1} \hspace{-0.1cm}- \hspace{-0.1cm}\gamma_j \sqrt{\mathrm{P}^a_{j,k-1}} \right], 
\label{eq:sigmpoints}
\end{eqnarray}
\vspace{-0.1cm}
\hspace{6cm} with $j\in\mathcal{N}_i$
} \vspace{-0.2cm}\\
Time-update: \vspace{-0.2cm}\\
\vbox{
\begin{eqnarray}
\hspace{-0.3cm}\mathcal{X}^{\mathrm{x}}_{i,k|k-1} &\hspace{-0.3cm}= &\hspace{-0.3cm} 
\sum\limits_{j \in \mathcal{N}_i} A_{ij} \mathcal{X}^{\x}_{j,k-1} 
+ f_i\left(\mathrm{z}_{i,k-1},\mathcal{X}^{\mathrm{x}}_{i,k-1},\hat{\ya}_{i,k-1},\mathcal{X}^{v}_{i,k-1} \right) \label{eq:timeup1}\\
\hspace{-0.3cm}\hat{\mathrm{x}}^-_{i,k} &\hspace{-0.3cm}= &\hspace{-0.3cm}\sum_{l=0}^{2n_i} W_{l,i}^{(m)}\mathcal{X}^{\mathrm{x}}_{l,i,k|k-1} \label{eq:timeup2}\\
\hspace{-0.3cm}\mathrm{P}^-_{i,k} &\hspace{-0.3cm}= &\hspace{-0.3cm}\sum_{l=0}^{2n_i} W_{l,i}^{(c)} \left( \mathcal{X}^{\mathrm{x}}_{l,i,k|k-1} - \hat{\mathrm{x}}^-_{i,k} \right) \left( \mathcal{X}^{\mathrm{x}}_{l,i,k|k-1} - \hat{\mathrm{x}}^-_{i,k} \right)^{\mathrm{T}} \label{eq:timeup3} \\
\hspace{-0.3cm}\mathcal{Y}_{i,k|k-1} &\hspace{-0.3cm}= &\hspace{-0.3cm}h_i\left(\mathrm{z}_{i,k},\mathcal{X}^{\mathrm{x}}_{i,k|k-1},\mathcal{X}^{w}_{i,k-1} \right) \\
\hspace{-0.3cm}\hat{\mathrm{y}}_{i,k}^- &\hspace{-0.3cm}= &\hspace{-0.3cm}\sum_{l=0}^{2n_i} W_{l,i}^{(m)}\mathcal{Y}_{l,i,k|k-1} \label{eq:timeupend}
\end{eqnarray}
} \vspace{-0.2cm}\\
Measurement-update \vspace{-0.2cm}\\
\vbox{
\begin{eqnarray}
\hspace{-0.3cm}\mathrm{P}_{\mathrm{y},i,k} &\hspace{-0.3cm}= &\hspace{-0.3cm}\sum_{l=0}^{2n_i} W_{l,i}^{(c)} \left( \mathcal{Y}_{l,i,k|k-1} - \hat{\mathrm{y}}_{i,k}^- \right) \left( \mathcal{Y}_{l,i,k|k-1} - \hat{\mathrm{y}}_{i,k}^- \right)^{\mathrm{T}} \label{eq:measup1} \\
\hspace{-0.3cm}\mathrm{P}_{\mathrm{xy},i,k} &\hspace{-0.3cm}= &\hspace{-0.3cm}\sum_{l=0}^{2n_i} W_{l,i}^{(c)} \left( \mathcal{X}^{\mathrm{x}}_{l,i,k|k-1} - \hat{\mathrm{x}}^-_{i,k} \right) \left( \mathcal{Y}_{l,i,k|k-1} - \hat{\mathrm{y}}_{i,k}^- \right)^{\mathrm{T}} \\
\hspace{-0.3cm}\mathcal{K}_{i,k} &\hspace{-0.3cm}= &\hspace{-0.3cm}\mathrm{P}_{\mathrm{xy},i,k} \mathrm{P}^{-1}_{\mathrm{y},i,k} \label{eq:measupmid} \\
\hspace{-0.3cm} \hat{\mathrm{x}}_{i,k} &\hspace{-0.2cm}= &\hspace{-0.2cm}\hat{\mathrm{x}}^-_{i,k} + \mathcal{K}_{i,k} \left( \mathrm{y}_{i,k} - \hat{\mathrm{y}}_{i,k}^- \right) 
\label{eq:stateestim} \\
\hspace{-0.3cm}\mathrm{P}_{i,k} &\hspace{-0.3cm}= &\hspace{-0.3cm}\mathrm{P}^-_{i,k} - \mathcal{K}_{i,k} \mathrm{P}_{\mathrm{y},i,k} \mathcal{K}_{i,k}^{\mathrm{T}} \label{eq:covestim}
\label{eq:covestim}
\end{eqnarray}
} \vspace{-0.2cm}\\
Parameters \vspace{-0.2cm}\\
\vbox{
\begin{equation}
\begin{array}{ll}
\hspace{-2.5cm} &\gamma_i = \sqrt{n_i + \lambda_i}, \hspace{0.3cm} \lambda_i = \alpha_i^2(n_i+\kappa_i)-n_i \\
\hspace{-2.5cm} &W^{(m)}_{0,i} = \frac{\lambda_i}{n_i + \lambda_i}, \hspace{0.3cm} W^{(c)}_{0,i} = \frac{\lambda_i}{n_i + \lambda_i} + 1 - \alpha_i^2 + \beta_i \\
\hspace{-2.5cm} &W^{(m)}_{l,i} = W^{(c)}_{l,i} = \frac{1}{2 (n_i + \lambda_i)}, \hspace{0.3cm} l = 1,\ldots,2n_i
\end{array}
\end{equation}
} \vspace{-0.2cm}\\ 
\hline
\textsuperscript{\textdagger}For compactness, the time argument is set as an index.
\end{tabular}
}


\label{tab:ukf}
\end{table}

Algorithm 1 states the proposed PUKF.
This PUKF can be interpreted as considering coupling states as inputs \cite{Khan-2008}, in contrast with e.g. state augmentation due to unknown input \cite{Singh-2014}. 
A similar strategy was followed in \cite{Minot-2016}, but the coupling between neighbors arose only in the measurement equations at each node. 

\begin{table}[!htb]
\centering


{\small
\begin{tabular}{l}
\hline
{\bf Algorithm 1} Distributed unscented Kalman filter algorithm \\ \hline
At each time step $k \geq 1$, subsystem $i$ \\
\forceindent 1. Measure $\mathrm{y}_i(k)$. \\
\forceindent 2. Broadcast to its neighbors the information set \\ 
\forceindent $\{ \hat{\x}_i(k), \Pa_{\x,i}(k) \}$. \\
\forceindent 3. Gather from its neighbors the information set \\
\forceindent $\{ \hat{\x}_j(k), \Pa_{\x,j}(k); j \in \mathcal{N}_i \}$. \\
\forceindent 4. Compute sigma-points $\mathcal{X}^a_j(k)$, $j \in \mathcal{N}_i$ using Eq. (\ref{eq:sigmpoints}). \\
\forceindent 5. Perform time-update using Eqs. (\ref{eq:timeup1})-(\ref{eq:timeupend}) and \\
\forceindent measurement-update using Eqs. (\ref{eq:measup1})-(\ref{eq:covestim}) to determine \\
\forceindent $\hat{\mathrm{x}}_i(k+1)$ and $\Pa_i(k+1)$.\\
\\ \hline
\end{tabular}
}


\label{tab:algo}
\end{table}




\section{SIMULATION STUDIES}	\label{sec:sim}

\begin{figure}[!htb]
\begin{center}
\includegraphics[scale=0.5]{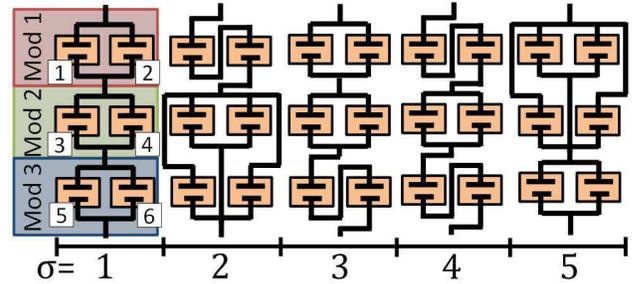}
\end{center}
\vspace{-0.8cm}
\caption{Battery pack representation of six battery cells electrically interconnected and subjected to reconfigurations $\sigma = \{1,\ldots,5 \}$.}
\label{fig:batcon}
\end{figure}

To assess the performance of the proposed PUKF, 
a small battery pack of $M = 6$ battery cells is considered. 
The battery pack layout is shown in Figure \ref{fig:batcon}. 
This deployment implies that battery cells $2i-1$ and $2i$, with $i = 1,\ldots,M$, are physically in parallel. 
Each pair of cells $(2i-1,2i)$ is grouped together to form a module. 
As seen below, the cells within a module behave similarly, which eases the presentation of the results. 
In the following, the simulation framework is firstly introduced and the results are explained next. 
The PUKF estimation is compared with that obtained from the centralized unscented Kalman filter (CUKF) and three PUKF versions arising from neglecting some model dynamics, namely electrolyte $\mathrm{C}_{e,i}(k) = [\mathrm{C}_{e1,i}(k), \mathrm{C}_{e2,i}(k)] = \mathrm{C}_{e,0}$, thermal coupling $A_{ij} = 0$ and temperature dynamics $\T_i(k) = [\T_{s,i}(k), \T_{c,i}(k)] = T_{\infty}$. 
While the comparison against the CUKF evidences the effectiveness of the proposed distributed UKF, the comparison against PUKF versions that neglect dynamics illustrates the importance of accounting for such dynamics.

%
%
%
%
%

\begin{figure}[h]
\begin{center}
\includegraphics[scale=0.48]{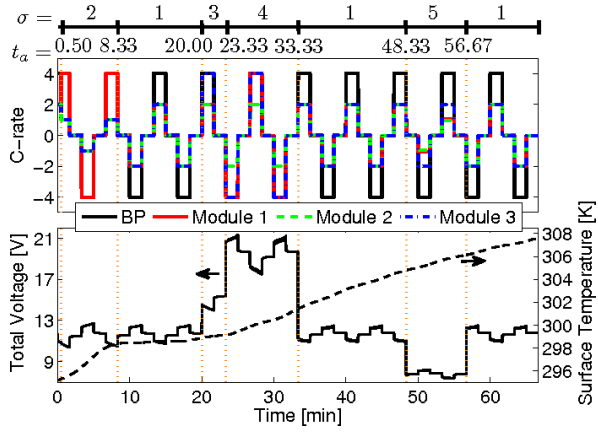}
\end{center}
\vspace{-0.3cm}
\caption{Gathered measurements from noise free simulation, namely upper plot: consecutive charge/discharge current pulses, lower plot: left $y$-axis, battery pack voltage and right $y$-axis, average surface temperature.}
\label{fig:totalsignals}
\end{figure}

Realistic virtual data of current, voltage and surface temperature has been gathered from the DFN-based battery pack simulator.
A total of 5 out of $2^5$ feasible electrical interconnection topologies have been arbitrarily chosen here for demonstration purposes. 
Each topology is depicted in Figure \ref{fig:batcon}, with the associated switching signal values $\sigma \in \{ 1,\ldots,5 \}$, and the time intervals $t_a$ when a given configuration is adopted are shown in the upper part of Figure \ref{fig:totalsignals}. 
The initial configuration $\sigma(0) = 1$ is the default one. It corresponds to a compromise between battery pack voltage and capacity. The other configurations result from electrical reconfigurations that might be useful to perform active balancing with reduced energy losses while coping with load demands. 
However the reconfiguration strategy is outside the scope of this paper.

The battery pack simulator was fed with consecutive charge/discharge current pulses of $4C$ (black solid curve in the upper plot of Figure \ref{fig:totalsignals}) spanning 40\% SOC, which corresponds to operating conditions close to plug-in hybrid electric vehicle applications. 
Nevertheless, in-pack battery cells might be subjected to different local input current magnitudes, always equal or proportionally less than the battery pack input current due to the reconfiguration capabilities of the considered battery (see colored curves in the upper plot of Figure \ref{fig:totalsignals}). 
The resulting battery pack output voltage and average surface temperature are shown in the lower plot, left and right $y$-axis of Figure \ref{fig:totalsignals}, respectively. 
The measurement noise in Eq. \eqref{eq:distout} corresponds to 
Gaussian noise sequences with covariance matrix given by $R_i = \mathrm{diag} \{ [ R_{11} \ R_{22} ] \}$ where $R_{11} = 10$ mV$^2$ and $R_{22} = 10^2$ mK$^2$.

 The PUKF was tuned with the following parameters
 
\vspace{-0.3cm}
{\small
\begin{equation}
\begin{array}{l}
\hspace{-0.4cm} \hat{\mathrm{x}}_{i,0} = \left[ [0.64, 0.64]\times10^{-2}, \left[1.05, 1.05\right] \times10^{3}, [295, 295] \right]^{\mathrm{T}} \hspace{-0.4cm} \\
\hspace{-0.4cm} \mathrm{diag}(\mathrm{P}_{i,0}) = 1_{n_i}^{\mathrm{T}} \times 10^{-8} \hspace{-0.4cm} \\
\hspace{-0.4cm} \mathrm{diag}(Q_i) = \left[ 0.1, 1_{m_i -1} \right]^{\mathrm{T}} \times 10^{-9} \hspace{-0.4cm} \\
\hspace{-0.4cm} \alpha_i = \alpha \sqrt{n_i} = 0.0245, \ \beta_i = 2, \ \kappa_i=3-n_i = -3 \	\label{eq:tuningparams}
\end{array}
\end{equation}
}
\vspace{-0.3cm}

\begin{figure*}[!htb]
\begin{center}
\includegraphics[scale=0.50]{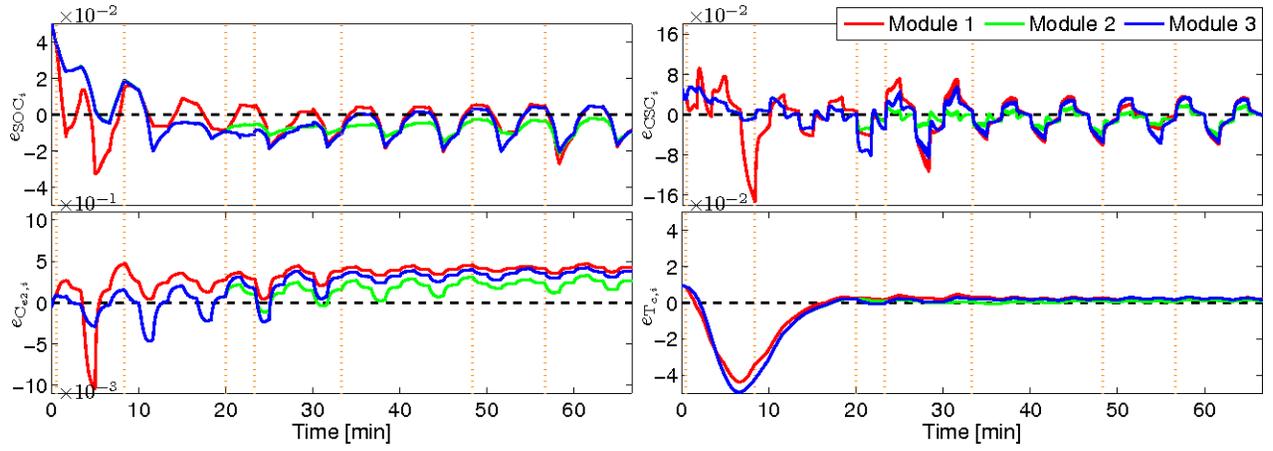}
\end{center}
\vspace{-0.3cm}
\caption{State estimation errors of the proposed PUKF for battery modules 1-3, namely upper plots: left, SOC and right, CSC, and lower plots: left, electrolyte concentration and right, core temperature.}
\vspace{-0.5cm}
\label{fig:PUKFstates}
\end{figure*}

where $\alpha_i$ was derived from $\alpha$ in order to preserve the consistency property, with $\alpha$ as the tuning parameter of the centralized UKF presented below. 
Notice that the related state variables, namely solid-phase diffusion $\left[ \mathrm{SOC}_i, \mathrm{CSC}_i \right]$, electrolyte diffusion $\left[ \mathrm{C}_{e1,i}, \mathrm{C}_{e2,i} \right]$ and thermal processes $\left[ \mathrm{T}_{c,i}, \mathrm{T}_{s,i}\right]$ share the same initialization due to the equilibrium assumption, i.e. battery cells are considered to be in a relaxed state at time zero ($t = 0$ min). Such initialization is valid for open circuit conditions (zero input current) during a long enough period of time (e.g. 1 hour). 
 Moreover, the initial state estimation error corresponds to 5\% for the unmeasurable state variables, namely $\left[ \widehat{\mathrm{SOC}}_i, \widehat{\mathrm{CSC}}_i, \hat{\mathrm{C}}_{e1,i}, \hat{\mathrm{C}}_{e2,i} \right]$, while the measurable surface temperature and its associated core temperature variable $\left[ \hat{\mathrm{T}}_{s,i}, \hat{\mathrm{T}}_{c,i} \right]$ were initialized with a 1\% error. The measurement noise covariance matrix was set to its actual value.
 
The PUKF estimation performance is studied in terms of the relative estimation error on the states defined as 
\vspace{-0.1cm}
$$e_{\mathrm{x}_i}(k) = \frac{ \mathrm{x}_i(k) - \hat{\mathrm{x}}_i(k) }{\mathrm{x}_i(k)}$$ 
Figure \ref{fig:PUKFstates} shows the obtained results, where 
the upper left and right plots correspond to the $\mathrm{SOC}$ and $\mathrm{CSC}$, respectively, while the lower left and right plots provide the $\mathrm{C}_{e2}$ and $\mathrm{T}_c$, respectively. 
Since all the in-pack battery cells were equally parameterized and initialized, and since the reconfiguration is performed in a 2-by-2 basis according to the adopted cell grouping, 
the state variables of the in-module battery cells are equivalent. Therefore, 
the estimation error of modules 1-3 are represented by the battery cells 1, 3 and 5 in Figure \ref{fig:PUKFstates} with curves color code as red, green and blue, respectively.

At the very beginning $\sigma(k) = 1$ with $k \in [0,0.50]$, time period during which 
the state estimation for the three modules is the same. 
Next, since $\sigma(k) = 2$ with $k \in (0.50,8.33]$ and until $t = 20$ min 
the states estimation of modules 2 and 3 are similar (green and blue curves are overlapped), 
whereas the state estimates of module 1 follow a different trajectory (red curves). 
After this second configuration, $\sigma(k) = 3$ with $k \in (20.00,33.33]$ and $\sigma(k) = 4$ with $k \in (23.33,33.33]$ are adopted. 
The reconfiguration of module 3 triggers the divergence between its states estimation (blue curves) and module 2 (green curves) while increasing the estimation error of this third module due to transient. 
A similar behaviour is portrayed by the estimation error of module 1 once the fourth configuration is adopted. 
Finally, $\sigma(k) = 5$ with $k \in (48.33,56.67]$ takes place.
Notice that module 2 reconfigurations make it to never experience the largest current ($4C$), which translates into the smallest estimation error for this module for most states and for most of the time.
Therefore, it can be concluded that the higher the current rate for a given battery cell the larger the state estimation error due to model mismatch.
Overall, the estimation error of the states oscilates around zero following the periodic trend of the input current pulses (upper plot of Figure \ref{fig:totalsignals}).

The PUKF is now compared with the estimation performance of the CUKF and the three considered versions of the PUKFs. 
The performance metric is the average root-mean-square error for the state estimation defined as
\vspace{-0.1cm}
$$\varepsilon_{\mathrm{x}} = \sqrt{ \frac{1}{N_t M} \sum\limits_{k}^{N_t} \sum\limits_{i}^M \left( \mathrm{x}_i(k) - \widehat{\mathrm{x}}_i(k) \right)^2 }$$
The tuning parameters of the CUKF are the centralized equivalent of Eq. (\ref{eq:tuningparams}), i.e. $\hat{{\bf x}}_0 = \mathrm{col}\{ \hat{\mathrm{x}}_{1,0},\ldots,\hat{\mathrm{x}}_{M,0} \}$, ${\bf P}_0 = \mathrm{diag}\{ \mathrm{P}_{1,0},\ldots,\mathrm{P}_{M,0} \}$ and ${\bf Q} = \mathrm{diag}\{ Q_1,\ldots,Q_M \}$, with $\alpha = 10^{-2}$, $\beta = \beta_i$ and $\kappa = \kappa_i$.
Table \ref{tab:RMSE} shows the obtained results normalized with respect to the considered CUKF denoted as $\bar{\varepsilon}_{\mathrm{x}}$.
Values greater than one in the table imply that the CUKF outperforms the correspondent filter for a given state estimate. 
From the table follows how a worse estimation of a given state may be countered by a better estimation of another state for the same filter with respect to the CUKF. 
The CUKF is slightly more accurate than the PUKF, which is expected. 
Nonetheless, the algorithm execution time at each sensor node within the PUKF is in average $10$ times smaller than the execution time of CUKF.
Although some states might be estimated more accurately by neglecting some dynamics (up to 97\% with respect to the CUKF), 
the states with largest errors can reach values up to 4\%, 8\% and 1137\% for the filters neglecting electrolyte, thermal coupling and temperature dynamics, respectively. 
This emphasizes the importance of accounting for extra dynamics and coupling when dealing with a battery pack subjected to current rates $\geq 1C$.

\begin{table}[!htb]
\caption{Normalized performance metric $\bar{\varepsilon}_{\mathrm{x}}$ 
for the PUKF and its considered variations.}
\vspace{-0.3cm}
\begin{center}
\begin{tabular}{ccccc}
\hline
$\bar{\varepsilon}_{\mathrm{x}}$ & PUKF &\hspace{-0.3cm} $\mathrm{C}_{e,i}(k) = \mathrm{C}_{e,0}$ &\hspace{-0.3cm} $A_{ij} = 0$ &\hspace{-0.3cm} $\T_i(k) = T_{\infty}$ \\ \hline
$\mathrm{SOC}$& $1.00$ &\hspace{-0.3cm} $1.04$ &\hspace{-0.3cm} $0.99$ &\hspace{-0.3cm} $1.20$ \\
$\mathrm{CSC}$& $0.99$ &\hspace{-0.3cm} $1.00$ &\hspace{-0.3cm} $0.98$ &\hspace{-0.3cm} $0.99$ \\
$\mathrm{C}_{e2}$& $1.00$ &\hspace{-0.3cm} $0.99$ &\hspace{-0.3cm} $1.00$ &\hspace{-0.3cm} $0.97$ \\
$\mathrm{T}_c$ & $1.00$ &\hspace{-0.3cm} $0.99$ &\hspace{-0.3cm} $1.08$ &\hspace{-0.3cm} $4.75$ \\
$\mathrm{T}_s$ & $1.00$ &\hspace{-0.3cm} $1.00$ &\hspace{-0.3cm} $0.99$ &\hspace{-0.3cm} $11.37$ \\
\hline
\end{tabular}
\end{center}
\label{tab:RMSE}
\end{table}

\section{CONCLUSIONS}

A partition-based distributed scheme for the state estimation of a lithium-ion battery pack has been presented. 
It amounts to place a sensor node per battery cell, to equip each node with a local unscented Kalman filter designed from an \emph{enhanced} reduced-order electrochemical model, and to allow the nodes to share suitable information. 
A simulated battery pack was subjected to dynamic reconfiguration scenarios, which impose different local currents on in-pack cells. 
The state estimation error increases with increments of the current magnitude. 
The distributed approach is able to provide state estimates as accurate as the centralized counterpart, but the algorithm execution time at each sensor node of the former approach is in average $10$ times smaller than for the latter approach. 
Model simplifications were shown to yield maximum errors between 4\% and 1137\% for given state variables. 
Ongoing work is devoted to exploit the obtained state estimates in order to control the battery pack reconfiguration to perform balancing.






\section*{ACKNOWLEDGMENT}

The authors would like to thank Professor Scott Moura for providing the battery simulation software that he developed.

\section*{APPENDIX}



The functions that characterizes the above presented model are given in Table \ref{tab:functions} for the sake of completeness (notice that subindex $i$ that characterizes each battery cell has been dropped for convenience).

\begin{table}[h]
\caption{Functions associated to the above introduced model.}
\centering
\begin{tabular}{c | c c}
\hline
Variable& Function &\spt Eq. involved\\ \hline
$\Phi^a$ & $\Phi_{\mathrm{ref}} \exp \left( \frac{E_{\Phi}}{R_g} \left( \frac{1}{T_{\mathrm{ref}}} - \frac{1}{\T_c(k)} \right) \right)$ &\spt (\ref{eq:statemat}),(\ref{eq:outputi}) \\ 
$D_{e,\mathrm{ref}}$ & $5.34\times 10^{-10} \exp \left( \frac{-0.65 \mathrm{C}_e(k)}{10^3} \right)$ 
&\spt (\ref{eq:statemat}) \\ 
$\Psi_{\mathrm{eff}} \ ^{b}$ & $\Psi \varepsilon_e^{\epsilon}$ &\spt (\ref{eq:statemat}) \\ 
$U_b^{\pm}$ & $\{ U_b^+ \left( \mathrm{SOC}^+ \right), U_b^- \left( \mathrm{SOC}(k) \right) \}^c$ &\spt (\ref{eq:Bm}) \\
$\frac{\partial{U_b^{\pm}}}{\partial{\mathrm{T}_{c}}}$ & $\left\{ \frac{\partial{U_b^+}}{\partial{\mathrm{T}_{c}}} \left( \mathrm{SOC}^+ \right), \frac{\partial{U_b^-}}{\partial{\mathrm{T}_{c}}} \left( \mathrm{SOC}(k) \right) \right\}^c$ &\spt (\ref{eq:Bm})\\ 
$U_s^{\pm}$ & $\{ U_s^{+} \left( \mathrm{CSC}^+ \right), U_s^{-}\left( \mathrm{CSC}(k) \right) \}^c$ &\spt (\ref{eq:outputi})\\
$\mathrm{SOC}^+$ & $\rho \mathrm{SOC}(k) + \sigma$ &\spt (\ref{eq:Bm}),(\ref{eq:outputi})\\
$\mathrm{CSC}^+$ & $\rho \mathrm{CSC}(k) + \sigma$ &\spt (\ref{eq:Bm}),(\ref{eq:outputi}) \\
$\rho, \sigma$ & $\rho = \frac{R_s^- L^- a_s^-}{c_{s,\mathrm{max}}^+ R_s^+ L^+ a_s^+}$, $\sigma = \frac{3 n_s^{Li}}{c_{s,\mathrm{max}}^+ R_s^+ L^+ a_s^+}$ &\spt (\ref{eq:Bm}),(\ref{eq:outputi}) \\
$\mathrm{C}_e^+$ & $\rho_e \mathrm{C}_e(k) + \sigma_e$ &\spt (\ref{eq:outputi}) \\
$\rho_e, \sigma_e$ & $\rho_e = -\frac{\varepsilon_e^- L^-}{\varepsilon_e^+ L^+}$, $\sigma_e = -\frac{\varepsilon_e^s L^s}{\varepsilon_e^+ L^+} \mathrm{C}_e^0 + \frac{n_e^{Li}}{\varepsilon_e^+ L^+ A}$ &\spt (\ref{eq:outputi}) \\
$\eta_s^{+}$ & $\frac{R_g \mathrm{T}_c(k)}{\alpha_{0} F} \sinh^{-1} \left( \frac{-1}{2 a_s^+ L^+ j^+_{n,0}} \mathrm{J}(k) \right)$ &\spt (\ref{eq:outputi}) \\
$\eta_s^{-}$ & $\frac{R_g \mathrm{T}_c(k)}{\alpha_{0} F} \sinh^{-1} \left( \frac{1}{2 a_s^- L^- j^-_{n,0}} \mathrm{J}(k) \right)$ &\spt (\ref{eq:outputi}) \\
$j^{\pm}_{n,0}$ & $k_{n}^{\pm} c_{s,\mathrm{max}}^{\pm} \sqrt{\mathrm{C}_{e}^{\pm}(k) \mathrm{CSC}^{\pm}(k) \left( 1 - \mathrm{CSC}^{\pm}(k) \right)}$ &\spt (\ref{eq:outputi}) \\
$\Delta \phi_e$ & 
\begin{tabular}{l}
\hspace{-0.3cm} $\frac{2 R_g \T_c(k)}{F\mathrm{C}_e^0} (  1-t_c^+ ) \left(  \mathrm{C}_e^+(k) - \mathrm{C}_e(k) \right) $ \\
\hspace{0.3cm} $- \frac{1}{\kappa} \left( \frac{L^+}{2 (\varepsilon_e^+)^\epsilon} + \frac{L^s}{(\varepsilon_e^s)^\epsilon} + \frac{L^-}{2 (\varepsilon_e^-)^\epsilon } \right) \mathrm{J}(k)$ 
\end{tabular}
&\spt (\ref{eq:outputi}) \\
\hline
\end{tabular}
\begin{tabular}{ll} 
\hspace{-1.2cm} $^a$$\Phi$ could be $D_s$, $D_e$, $k_n$ or $\kappa$ & $^b$$\Psi$ could be $D_e$ or $\kappa$ \\
\hspace{-1.2cm} $^c$Taken from \cite{Mao-2014} \\
\end{tabular}
\label{tab:functions}
\vspace{-0.3cm}
\end{table}

Nomenclature for Table \ref{tab:functions} is introduced in Table \ref{tab:nom}.
\begin{table}[!h]
\caption{Nomenclature for Table \ref{tab:functions}.}
\centering

{\normalsize
\begin{tabular}{l l}
%
\hline
Symbol&\hspace{-0.2cm} Parameter \\ 
\hline
$\Phi_{\mathrm{ref}}$ &\hspace{-0.2cm} Variable $\Phi$ at the reference temperature \\
$E_{\Phi}$ &\hspace{-0.2cm} Activation energy of variable $\Phi$ (J.mol$^{-1}$) \\
$R_g$ &\hspace{-0.2cm} Universal gas constant (=8.31 J.mol$^{-1}$K$^{-1}$) \\
$T_{\mathrm{ref}}$ &\hspace{-0.2cm} Reference temperature (K) \\
$D_e$ &\hspace{-0.2cm} Electrolyte diffusion coefficient (m$^2$.s$^{-1}$) \\
$L$ &\hspace{-0.2cm} Electrode/separator thickness (m) \\
$\varepsilon_e$ &\hspace{-0.2cm} Electrolyte volume fraction \\
$t_c^+$ &\hspace{-0.2cm} Transference number \\
$A$ &\hspace{-0.2cm} Cross-sectional battery cell area (m$^2$) \\
$\epsilon$ &\hspace{-0.2cm} Bruggeman's exponent \\
$n^{Li}$ &\hspace{-0.2cm} Total amount of lithium (mol) \\
$\alpha_0$ &\hspace{-0.2cm} Apparent transfer coefficient \\
$k_n$ &\hspace{-0.2cm} Reaction rate constant (A.m$^{2.5}$.mol$^{-1.5}$) \\ \hline
\end{tabular} \\
}


\label{tab:nom}
\vspace{-0.3cm}
\end{table}

In order to derive an electrolyte \emph{enhanced} reduced-order EChM, 
the approach proposed in \cite{Rahn-2013ch4} is leveraged to analytically solve the electrolyte diffusion PDEs given by
\begin{equation}
\varepsilon_e \frac{\p c_e}{\p t}(x,t) = D_{e,\mathrm{eff}} \frac{\p^2 c_e}{\p x^2}(x,t) + a_s (1-t_c^+) j_n(x,t) \label{eq:ce}
\end{equation}

where $c_e$ is the electrolyte concentration that covers the entire battery cell thickness. Eq. \eqref{eq:ce} takes the stated form within the positive $(+)$ and negative $(-)$ electrode spatial domains, while $\varepsilon_e = 1$ and $j_n = 0$ within the separator $(s)$ domain. 
The boundary conditions that guarantee a zero flux of lithium ions outside the system and continuity of ion flux and electrolyte concentration throughout the cell thickness are given by
\begin{subnumcases}{\label{eq:cebcs}\hspace{-0.8cm}} 
\hspace{-0.2cm} \left.\frac{\partial c_{e}^{-}}{\partial x}(x,t)\right|_{x=0} = \left.\frac{\partial c_{e}^{+}}{\partial x}(x,t)\right|_{x=L}=0\label{eq:cebc1} \\ 
\hspace{-0.2cm} \left. D_{e,\mathrm{eff}}^{-}\frac{\partial c_{e}^{-}}{\partial x}(x,t)\right|_{x=L_{n}} = \left. D_{e,\mathrm{eff}}^{s} \frac{\partial c_{e}^{s}}{\partial x}(x,t)\right|_{x=L_{n}} \label{eq:cebc2} \\ 
\hspace{-0.2cm} \left. c_{e}^{-}(x,t)\right|_{x=L_{n}}=\left. c_{e}^{s}(x,t)\right|_{x=L_{n}}\label{eq:cebc3} \\ 
\hspace{-0.2cm} \left. D_{e,\mathrm{eff}}^{s} \frac{\partial c_{e}^{s}}{\partial x}(x,t)\right|_{x=L_{ns}}=\left. D_{e,\mathrm{eff}}^{+}\frac{\partial c_{e}^{+}}{\partial x}(x,t)\right|_{x=L_{ns}} \label{eq:cebc4} \\ 
\hspace{-0.2cm} \left. c_{e}^{s}(x,t)\right|_{x=L_{ns}}=\left. c_{e}^{+}(x,t)\right|_{x=L_{ns}} \label{eq:cebc5} 
\end{subnumcases}

Expanding Eq. \eqref{eq:ce} with superindices to denote spatial domains and taking the Laplace transform to eliminate the time derivative results in
\begin{equation}
s \varepsilon_e^{\pm s} \mathrm{C}_e^{\pm s}(x,s) - D_{e,\mathrm{eff}}^{\pm s} \frac{\da^2 \mathrm{C}_e^{\pm s}}{\da x^2}(x,s) + b^{\pm s} \mathrm{J}(s) = 0 \label{eq:celt}
\end{equation}

where $b^{\pm} = \mp \frac{1-t_c^+}{F L^{\pm} A}$ and $b^{s} = 0$, the pore-wall molar flux has been replaced by the uniform utilization 
$j_n(x,t) \approx \frac{\mathrm{J}(t)}{F a_s L }$, 
$\mathrm{C}_{e}(x,s)$ and $\mathrm{J}(s)$ are the Laplace transforms of $c_{e}(x,t)$ and $\mathrm{J}(t)$ respectively, with $s$ as the Laplace variable. 
The solution of Eq. \eqref{eq:celt} is
\begin{equation}
\mathrm{C}_e^{\pm s}(x,s) = K_1^{\pm s} \exp( \beta^{\pm s} x ) + K_2^{\pm s} \exp( -\beta^{\pm s} x ) + \frac{b^{\pm s} }{\varepsilon_e^{\pm s} s}\mathrm{J}(s) \label{eq:cesol}
\end{equation}

where $\beta^{\pm s} = \sqrt{\frac{\varepsilon_e^{\pm s} s}{D_e^{\pm s}}}$. 
Substituting Eq. \eqref{eq:cesol} into boundary conditions Eqs. \eqref{eq:cebcs} produces six linear equations with unknown constants $\{ K_1^+, K_2^+, K_1^-, K_2^-, K_1^s, K_2^s \}$. After solving such system and evaluating it at the current collector/negative electrode interface $x = 0$, the transcendental transfer function of the form
\begin{equation}
\frac{\mathrm{C}_e (s)}{\mathrm{J} (s)} = \frac{N_c (s)}{D_c(s)} \label{eq:cetf}
\end{equation}

with 
\begin{equation}
\begin{array}{lcl}
N_c (s) \hspace{-0.2cm}&= 
\hspace{-0.2cm}&b^-\left( \alpha^-\varepsilon_e^+\sinh(\beta^- L^-)\sinh(\beta^+ L^+)\sinh(\beta^s L^s) \right. \\
&&\hspace{-0.8cm}	+ \alpha^s\varepsilon_e^+\cosh(\beta^- L^-)\cosh(\beta^s L^s)\sinh(\beta^+ L^+) \\
&&\hspace{-0.8cm}	+ \alpha^-\alpha^+\alpha^s D_e \cosh(\beta^+ L^+)\cosh(\beta^s L^s)\sinh(\beta^- L^-) \\
&&\hspace{-0.8cm}	+ \alpha^+\varepsilon_e^s\cosh(\beta^- L^-)\cosh(\beta^+ L^+)\sinh(\beta^s L^s) \\
&&\hspace{-0.8cm}	- \alpha^s\varepsilon_e^+\cosh(\beta^s L^s)\sinh(\beta^+ L^+) \\
&&\hspace{-0.8cm}	\left. - \alpha^+\varepsilon_e^s\cosh(\beta^+ L^+)\sinh(\beta^s L^s) \right) \\
&&\hspace{-0.8cm}	- b^+ \alpha^s \varepsilon_e^-\sinh(\beta^+ L^+)
\end{array}
\label{eq:cetf1}
\end{equation}
\begin{equation}
\begin{array}{lcl}
D_c(s) \hspace{-0.2cm}&= 
\hspace{-0.2cm}&\varepsilon_e^- s \left( \alpha^s \varepsilon_e^+\cosh(\beta^- L^-)\cosh(\beta^s L^s)\sinh(\beta^+ L^+)  \right. \\
&&\hspace{-0.8cm}	  + \alpha^+ \varepsilon_e^s\cosh(\beta^- L^-)\cosh(\beta^+ L^+)\sinh(\beta^s L^s) \\
&&\hspace{-0.8cm}	  + \alpha^- \varepsilon_e^+\sinh(\beta^- L^-)\sinh(\beta^+ L^+)\sinh(\beta^s L^s) \\
&&\hspace{-0.8cm}	  \left. + \alpha^-\alpha^+\alpha^s D_e \cosh(\beta^+L^+)\cosh(\beta^s L^s)\sinh(\beta^- L^-) \right)
\end{array}
\label{eq:cetf2}
\end{equation}

where $\beta^{\pm s}$ is as previously defined and $\alpha^{\pm s} = \sqrt{\frac{\varepsilon_e^{\pm s}}{D_e^{\pm s}}}$. 
Only the negative electrode external boundary $x = 0$ is taken to evaluate the electrolyte diffusion. 
Such electrode choice is in line with the reference electrode for solid-phase diffusion while the external boundary location is taken since it corresponds to one of the voltage terminals. 
The electrolyte concentration at the positive voltage terminal results from the material balance giving rise to a linear relationship $\mathrm{C}_e^+(k) = \rho_e \mathrm{C}_e(k) + \sigma_e$ with constants $\rho_e$ and $\sigma_e$ given in Table \ref{tab:functions}. 

The transfer function Eq. \eqref{eq:cetf} is truncated through a second-order Pad\' e approximation and parameterized as an equivalent-hydraulic model such as
\begin{equation}
\frac{\mathrm{C}_e (s)}{\mathrm{J} (s)} = \gamma_e \frac{\beta_e s + \mathrm{g}_e}{s ( \beta_e (1 - \beta_e) s + \mathrm{g}_e )} \label{eq:pade}
\end{equation}

which is finally rewritten in state-space form and discretized in time via Euler's approximation to get the following $e$EHM 
\begin{equation}
\mathrm{C}_e (k+1) = 
\left[ \hspace{-0.2cm}
\begin{array}{cc}
1 \hspace{-0.2cm}&0 \hspace{-0.2cm} \\
\frac{\mathrm{g}_{e}}{b_{e}} \hspace{-0.2cm}&1 - \frac{\mathrm{g}_{e}}{b_{e}}
\end{array}
\hspace{-0.2cm} \right]
\mathrm{C}_e (k) 
+
\left[ \hspace{-0.2cm} 
\begin{array}{c}
 -\gamma_e \\
\frac{\gamma_e}{1 - \beta_{e}} \\
\end{array}
\hspace{-0.2cm} \right] 
\mathrm{J}(k)
\end{equation}

with $b_e = \beta_e ( 1 - \beta_e )$ and the state vector $\mathrm{C}_e(k) = [\mathrm{C}_{e1}(k), \mathrm{C}_{e2}(k)]^\tr$.

The linearized electrolyte charge conservation equation is given by
\begin{equation}
\kappa_{\mathrm{eff}} \frac{\p^2 \phi_{e}}{\p x^2}(x,t) = \frac{2R_g\mathrm{T}(t)}{F \mathrm{C}_e^0}(1-t_{c}^{+}) \frac{\p^2 \mathrm{C}_e}{\p x^2}(x,t) - F a_s j_n(x,t)  \label{eq:phie}
\end{equation}

where $\phi_e$ is the electrolyte electric potential that covers the entire battery cell thickness. 
Since the voltage response results from the potential difference between battery terminals, $\phi_e$ can be set to zero at $x=0$ and only potential differences might be considered. Defining the electrolyte potential difference along the cell thickness as $\Delta \phi_e = \phi_e^+(L,t) - \phi_e^-(0,t)$ and integrating directly Eq. \eqref{eq:phie}, the following expression arises
\begin{equation}
\begin{array}{lcl}
\Delta \phi_e(t) \hspace{-0.2cm}&= \hspace{-0.2cm}&\frac{2R_g\mathrm{T}_c(t)}{F \mathrm{C}_e^0}(1-t_{c}^{+})(\mathrm{C}_e^+(t) - \mathrm{C}_e(t)) \\ 
&&\hspace{0.6cm} - \frac{1}{\kappa} \left( \frac{L^+}{2 (\varepsilon_e^+)^\epsilon} + \frac{L^s}{(\varepsilon_e^s)^\epsilon} + \frac{L^-}{2 (\varepsilon_e^-)^\epsilon } \right) \mathrm{J}(t) 
\label{eq:Deltphie}
\end{array}
\end{equation}

where the assumption of uniform utilization has been used. 
The continuous-time variable $t$ can be replaced by the discrete-time one $k$ given that Eq. \eqref{eq:Deltphie} is algebraic.


\bibliography{ACC2018.bbl}

\end{document}